# The Nexus of AR/VR, AI, UI/UX, and Robotics Technologies in Enhancing Learning and Social Interaction for Children with Autism Spectrum Disorders: A Systematic Review

**Biplov Paneru[a]**

[a]Department of Electronics and Communication Engineering, Nepal Engineering College, Pokhara University, Bhaktapur Nepal

corresponding author: Biplov Paneru, email: biplov001@gmail.com

*Abstract*

The emergence of large language models (LLMs), augmented reality (AR), and user interface/user experience (UI/UX) design in therapies for children, especially with disorders like autism spectrum disorder (ASD), is studied in detail in this review study. 150 publications were collected by a thorough literature search throughout PubMed, ACM, IEEE Xplore, Elsevier, and Google Scholar; 60 of them were chosen based on their methodological rigor and relevance to the focus area. Three of the primary areas are studied and covered in this review: how AR can improve social and learning results, how LLMs can support communication, and how UI/UX design affects how effective these technologies can be. Results show that while LLMs can provide individualized learning and communication support, AR has shown promise in enhancing social skills, motivation, and attention. For children with ASD, accessible and engaging interventions rely heavily on effective UI/UX design, but there is still a significant lack of robotics-based education and therapeutic programs specifically tailored for autistic children. To optimize the benefits of these technologies in ASD therapies and immersive education, the study emphasizes the need for additional research to address difficulties related to customization, accessibility, and integration.

1. **Introduction**

Autism spectrum disorder (ASD) is a neurodevelopmental illness that makes it hard for people to communicate and connect socially. It also causes people to have limited, repetitive habits, interests, or activities [19].People with autism spectrum disorders have serious and widespread problems with multiple areas of development, such as communication abilities, social skills, or having repetitive behaviors, hobbies, and activities [20]. Children with autism can benefit greatly from digitally assisted language therapies thanks to augmented reality (AR). Numerous results and insights about the use of augmented reality (AR) as a teaching and pedagogical aid have been reported by educators and researchers [1]. The use of computer technology—particularly augmented reality—in autism spectrum disorder (ASD) therapies has grown as a means of treating or mitigating the symptoms of the disorder. Not just for kids of a certain age or educational level, augmented reality is an entertaining form of technology that facilitates easy interaction and helps kids comprehend and retain information [2]. A neurodevelopmental disorder known as autism spectrum disorder (ASD) is marked by recurring problems with social interaction and communication, as well as a limitation in interests and repetitive activities [3]. It is believed that one in every 100 youngsters worldwide is affected by ASD. Developments in autism research have been greatly enhanced by notable breakthroughs in international policy [4]. VR and Emotional Games. AR significantly improves the interactiveness of the learning process for kids with ASD [5, 6, 7]. The rise of large language models (LLMs) such as ChatGPT has revolutionized information access, providing instant answers to a wide range of questions [13].

Emerging technologies like virtual reality (VR) and augmented reality (AR) are being used more frequently in research. This has made it possible to create interactive systems that are now a component of contemporary teaching techniques [40]. Autism Spectrum Disorder (ASD) affects children's ability to communicate, interact socially, and engage in learning activities. Traditional educational and therapeutic approaches often struggle to meet the unique needs of children with ASD, leading researchers and practitioners [21] to explore innovative technological solutions. Among these, Augmented Reality (AR), Large Language Models (LLMs), and effective User Interface/User Experience (UI/UX) design have emerged as promising tools to enhance learning and social interaction for children with ASD [10, 11].

## 1.1 Background of the study

Social contact, communication, and increased sensitivity to stimuli are among the main issues faced by children with autism spectrum disorder (ASD), underscoring the pressing need for creative intervention strategies [1, 2, 5]. ASD education and therapy have seen a growing use of technological advancements like augmented and virtual reality (AR/VR), artificial intelligence (AI), including large language models (LLMs), user interface/user experience design (UI/UX), robotics [1-10] and gaming systems for cognitive improvements [5, 49]. Robotic therapies are gaining increasing attention in the autism sector, particularly for enhancing the social skills of children along with various products like assistive technologies [15], supplementing traditional human interventions [60]. These advancements have marked significant turning points in improving engagement and providing individualized support. However, there is a noticeable paucity of long-term studies assessing these technologies' long-term efficacy, and current research frequently lacks integrated approaches that combine different technologies—for example, AR with LLMs. Many people believe that current diagnostic techniques for autistic spectrum disorder are ineffective. Developments in Human-Robot Interactions (HRI) offer promising new diagnostic techniques utilizing interactive robots [53]. Cognitive behavioral therapy too (CBT) is effective for managing multiple neurological and psychiatric disorders. It enhances anxiety symptoms in children with autism spectrum disorder, receiving substantial empirical backing [54]. AI-driven systems can supports social skill development through interactive simulations and promotes long-term societal benefits by helping autistic individuals contribute meaningfully across diverse fields [16]. Although these technologies offer considerable advantages, they also introduce distinct difficulties for the autistic and broader neurodivergent community. This analysis critically explores the effects of LLMs in this community, highlighting risks of misinformation, the reinforcement of stigmatizing medical models, the misrepresentation of non-speaking neurodivergent people, and challenges related to equity and privacy. Through individual experimentation with LLMs and conversations with community members, issues are highlighted regarding the reinforcement of stereotypes, the danger of excessive dependence on possibly erroneous AI results, and the insufficient support for varied communication requirements [13]. Together these systems can be described as:

1. Augmented Reality (AR): It is a technology that overlays digital images or information on the real world through devices like smartphones or glasses, enhancing what you see around you.
2. Virtual Reality (VR): This is a fully immersive digital environment created by computers, where users can interact in a simulated space using special headsets.
3. Artificial Intelligence (AI): They are computer systems designed to perform tasks that normally require human intelligence, such as learning, reasoning, and problem-solving.
4. Large Language Models (LLMs): They are the advanced AI models trained on vast amounts of text to understand and generate human-like language, enabling chatbots, language translation, and more.
5. User Interface/User Experience Design (UI/UX): UI refers to the layout and interactive elements of a software or app, while UX focuses on the overall experience and ease of use for the user.
6. Robotics: The design and use of robots, which are machines that can perform tasks autonomously or semi-autonomously.
7. Human-Robot Interaction (HRI): The study and development of how humans and robots communicate and work together effectively.
8. Cognitive Behavioral Therapy (CBT): A type of psychological treatment that helps people change negative thought patterns and behaviors to improve mental health.

This paper systematically reviews the role of these technologies in addressing the challenges associated with ASD, focusing on their effectiveness, limitations, and potential for future applications. There is lack of research work in the field of training the children with Autism Spectrum Disorder [40]. The term "autism spectrum disorder" (ASD) describes a broad range of neurodevelopmental disorders that affect how people interact with their environment. It includes a variety of issues with social interaction, interpersonal connections, and confined or repetitive hobbies and habits [41]. One of the most difficult facets of contemporary education is inclusive education. Discovering cutting-edge resources that could modify the educational process to meet the needs of students with disabilities has proven to be a significant obstacle in the recent fight for educational inclusion [43]. Virtual reality (VR) has the potential to produce socially realistic, perceptually rich media simulations including virtual characters that may have a distinct impact on young children's responses to content compared to traditional mediums like television [44]. Gamified approaches and serious games have been extensively utilized by therapists and researchers collaborating with individuals on the autism spectrum [55]. A study by Ramos Aguia et al., (2023) investigates gamification to enhance user engagement for individuals with disabilities by examining two case studies: an application that educates blind users about Mexican currency and an application that assists those with autism in navigating their surroundings. Findings indicated strong usability for the app designed for blind users and enhanced performance for users with autism. Although the results indicate that gamification holds potential for enhancing accessibility, the research is constrained by minimal sample sizes and insufficient information on the particular gamification methods employed. It still provides important initial insights and emphasizes the possibilities of gamification in assistive technologies [57].

Social and emotional learning (SEL) has drawn interest from all across the world in recent years. "The process by which all young people and adults acquire and apply the knowledge, skills, and attitudes to develop healthy identities, manage emotions and accomplish personal and collective goals, feel and show empathy for others, establish and maintain supportive relationships, and make responsible and caring decisions" is the definition of support-oriented learning (SEL) [45]. The emergence of large language models (LLMs) and their greater availability through chatbots such as OpenAI's ChatGPT (2023), Google's Bard (2023), and Microsoft's Bing Chat (2023) present educationalists with both new potential and problems [49].

## 2. Methodology

A comprehensive literature search was conducted across multiple databases, including PubMed, IEEE Xplore, Springer, and Google Scholar, using keywords such as "Augmented Reality," "Autism Spectrum Disorder," "Large Language Models," "UI/UX design," and "Children with ASD." The search yielded 150 articles, of which 25 were selected for detailed analysis based on their relevance to the integration of AR, LLM, and UI/UX in interventions for children with ASD. The selected studies were evaluated based on their methodological rigor, the clarity of intervention descriptions, and the reported outcomes.

### 2.1 Inclusion and Exclusion Criteria

1. Inclusion Criteria: Studies published between 2008 and 2025, focusing on children with ASD, using AR, LLM, or UI/UX interventions, and reporting measurable outcomes related to learning, social interaction, or emotional understanding.

2. Exclusion Criteria: Studies that did not focused on education and knowledge, social interactions behaviors and did not utilize AR, LLM, or UI/UX technologies, or lacked empirical evidence.

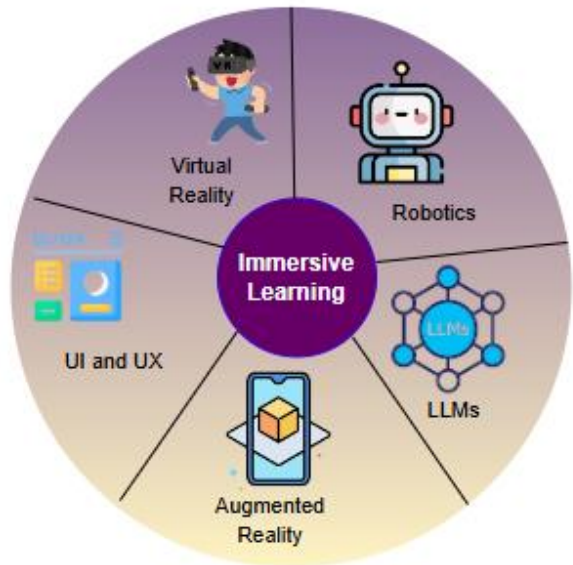

Fig 1. Technologies for immersive learning for children

2.2 Data Extraction

Data from the selected studies were extracted and categorized into three main themes: the role of AR in enhancing social and learning outcomes, the application of LLMs in ASD interventions, and the impact of UI/UX design on the effectiveness of these technologies. The extracted data included study objectives, sample characteristics, intervention details, and key findings.

**2.3 Block Diagram of Methodology**

Below is a block diagram shown in figure 2. The block diagram shows the methodology used in this systematic review:

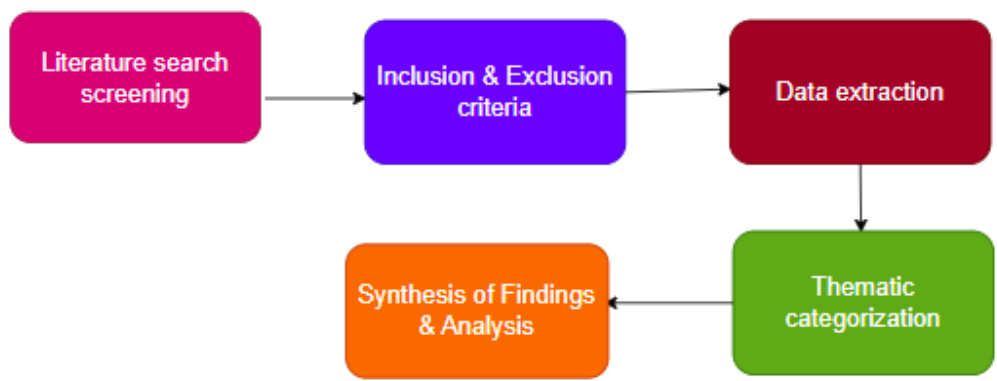

Fig 2. Overall block diagram

2.4 **Literature Search Strategy**

To ensure a comprehensive and unbiased collection of relevant studies, the following steps were taken:

- **Databases Searched**: PubMed, IEEE Xplore, ACM Digital Library, Elsevier, Springer, and Google Scholar.
- **Keywords Used**: Combinations of terms such as "Autism Spectrum Disorder," "Augmented Reality," "Virtual Reality," "Large Language Models," "UI/UX design," "Children with ASD," "Social Interaction," and "Learning Technologies."
- **Time Frame**: Studies published between 2008 and 2024 were included to capture the evolution of technological interventions.

2.5 **Inclusion and Exclusion Criteria**

**Inclusion Criteria**:

i. Peer-reviewed studies focusing on children with ASD.
ii. Interventions utilizing AR, VR, LLMs, or UI/UX design.
iii. Empirical studies with measurable outcomes (e.g., social skills, communication, engagement).
iv. Studies published in English.

**2.6 Exclusion Criteria**:

i. Studies not involving children with ASD.
ii. Non-technological interventions (e.g., traditional therapies without digital components).
iii. Lack of empirical data or methodological rigor.

2.7 **Study Selection Process**

i. Initial Screening: Titles and abstracts of 150 identified studies were screened for relevance.

ii. Full-Text Review: 80+ studies meeting the inclusion criteria underwent detailed evaluation.
iii. Final Selection:60 studies were chosen based on methodological quality, relevance, and contribution to the research objectives.

**2.8 Data Extraction and Categorization**

Data from selected studies were systematically extracted and categorized into three themes:

1. AR/VR Interventions: Focused on social skills, learning outcomes, and emotional regulation.
2. LLMs in Communication: Examined personalized learning, language support, and integration with AR/VR.
3. UI/UX Design: Analyzed interface adaptability, sensory considerations, and engagement metrics.

2.9 Quality Assessment

- Methodological Rigor: Studies were evaluated for clear research design, sample size, and validity of outcomes.
- Bias Mitigation: Only peer-reviewed studies with transparent methodologies were included.
- Inter-Rater Reliability: Two independent reviewers assessed study quality, with discrepancies resolved through discussion.

2.10 **Synthesis and Analysis**

i. Qualitative Synthesis: Thematic analysis of key findings, challenges, and trends.
ii. Quantitative Synthesis: Meta-analysis of effect sizes (e.g., Hedges' g) for studies with comparable outcomes (e.g., VR's impact on social skills).
iii. Visual Representation: Findings were summarized using tables (e.g., Table 1) and figures (e.g., Figures 8–10) to highlight trends and distributions.

**2.11 Limitations and Ethical Considerations**

i. Limitations: Heterogeneity in study designs and small sample sizes in some studies.

ii. Ethical Compliance: Ensured all included studies adhered to ethical guidelines for research involving children with ASD.

**2.12 Contextual Suitability**

This methodology aligns with the interdisciplinary nature of the review, bridging technology (AR/VR, LLMs, AI and robotics oriented systems) and human-centered design (UI/UX) to address ASD-specific challenges. By combining systematic data extraction with thematic and statistical analysis, it provides a robust framework for evaluating the efficacy and future potential of these interventions.

### 3. AR / VR and AI-based systems applications in ASD

3.1 Augmented Reality (AR) and virtual Reality (VR) in ASD Interventions

AR has been widely studied for its potential to enhance learning and social interaction in children with ASD. The immersive and interactive nature of AR allows children to engage in scenarios that mimic real-life social interactions, which can be challenging for those with ASD. Studies by Shemy et al. (2024) [1], Wedyan et al. (2021) [2], and Berenguer et al. (2020) [3] have demonstrated that AR can significantly improve children's understanding of facial expressions, social cues, and language skills by providing a safe and controlled environment for practice. While augmented reality is typically linked to software innovations, creating applications in the autism sector requires examining the relevant technology and focusing on the interaction elements with augmented reality tools. Augmented Reality is a method that can introduce social therapy into a virtual environment to enhance intrinsic motivation in children with Autism Spectrum Disorder (ASD) Adnan et al .(2018) [12].

Moreover, AR-based applications have been shown to increase attention span and motivation in children with ASD, which are critical factors in learning [4]. However, the effectiveness of AR interventions depends heavily on the customization of content to meet individual needs, as well as the involvement of caregivers and educators in the implementation process [5]. Additionally, Bhatt et al. (2014) [5] highlight the importance of integrating AR with other therapeutic approaches to maximize its benefits. An example user interface is shown in figure 3.

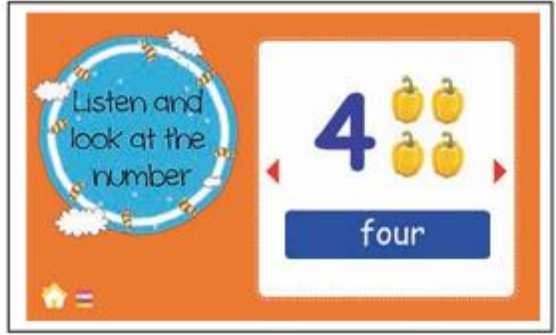

Fig 3. UI example for gaming focused on children [51]

In contrast to conventional emotional training carried out by therapists, this study examines the efficacy of virtual reality (VR) as an intervention method for enhancing social skills in children with autism spectrum disorder (ASD).
The main purpose of virtual reality in rehabilitation is to help people with disabilities practice their impaired functions so they can manage, conquer, lessen, or make up for their impairments [23].

The goal of the study is to determine which intervention takes the least amount of time to acquire for both primary and secondary emotions. Two groups of sixty kids with level 1 ASD were created, one for VR-based training and the other for conventional therapy. The VR group had a faster acquisition time for recognizing and employing both primary and secondary emotions, but both groups showed similar acquisition times for primary emotion recognition. According to the study, VR could be a fun and useful tool for helping kids with ASD strengthen their social skills. The VR intervention offers a high degree of involvement and the possibility for real-life skill application, which is in line with recent developments in the use of technology to support people with ASD. These results add to the increasing amount of data that supports the use of VR in therapeutic settings for people with ASD [31]. The figure 4 shows the experimental comparison carried out.

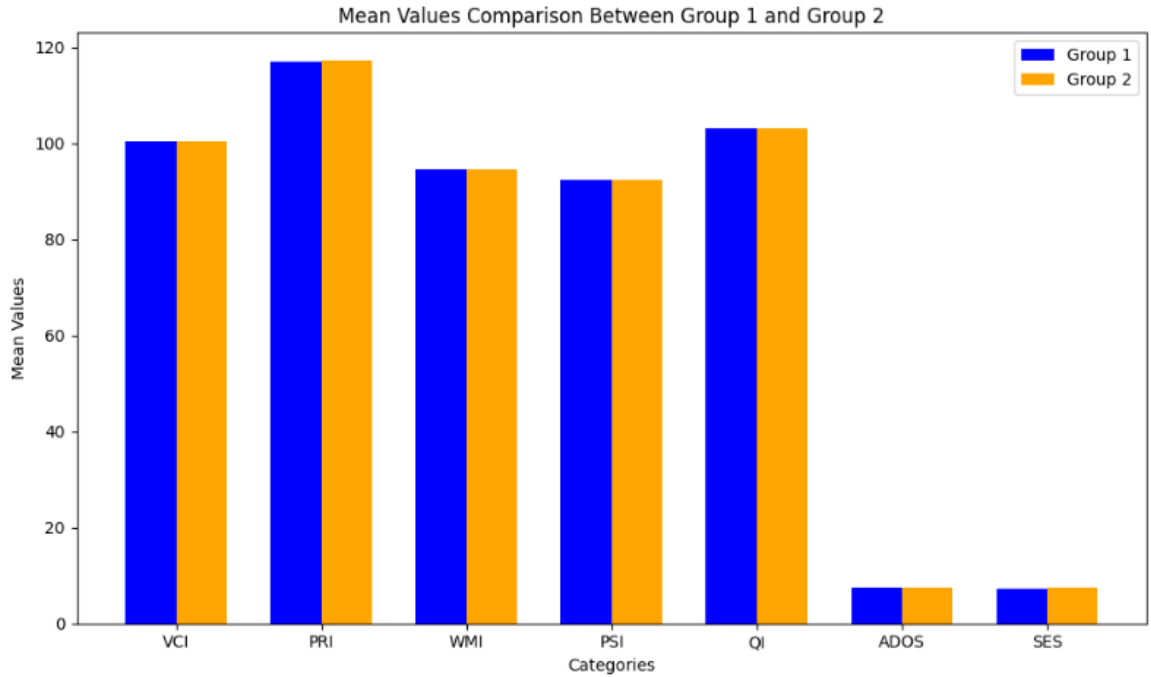

Fig 4. Inclusion criteria based comparison between involved groups [31]

The increasing use of virtual reality (VR) as a tool to meet the psychological requirements of people with high-functioning autism, particularly in the population with Autism Spectrum Disorder (ASD), is highlighted in this scoping review. The report groups data on social skills, eye gazing and joint attention, motor learning, and job training after analyzing 23 experimental trials. Notwithstanding the potential of virtual reality, the study reveals a paucity of studies providing useful therapies to improve adaptive functioning in people with high-functioning ASD. Notably, half of the studies included participants under the age of 18, indicating that, in order to optimize long-term effects, future research should concentrate on early phases of development. The study emphasizes the necessity of more focused therapies to help young individuals with ASD get ready for obstacles in life [33]. A review by Mesa-Gresa et al. (2018) finds moderate evidence supporting VR's effectiveness in improving ASD symptomatology, while emphasizing the need for more robust validation to confirm VR as a complement to traditional therapies. Overall, the paper offers a valuable and methodologically thorough contribution, advancing the understanding of VR's role in ASD treatment and highlighting gaps for future research [14].

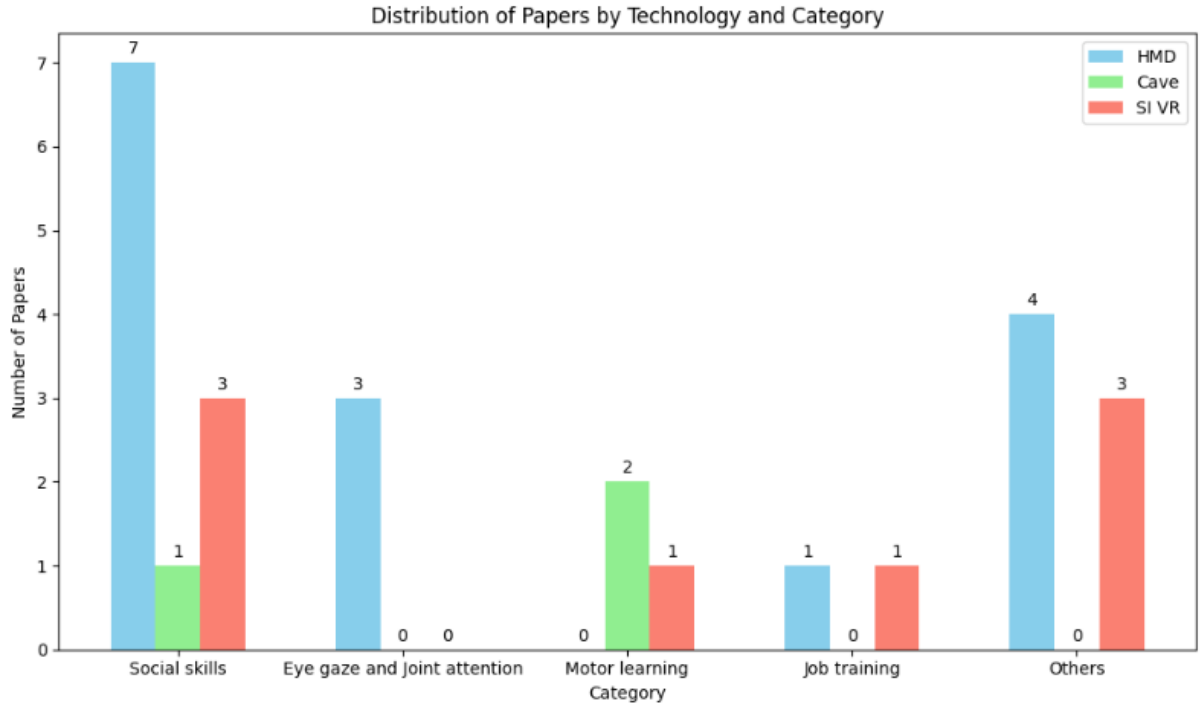

Fig 5. Distribution of papers considering the defined 5 categories (i.e., social interaction, eye gaze and joint attention, motor learning, job training, and others) [33]

VR is characterized as an immersive experience generated by computer-created settings. Its uses vary from basic technological configurations to more intricate immersive systems [24]. The usefulness of immersive virtual reality (VR) therapy in treating phobias in people with autism spectrum disorder (ASD) is investigated in this study. Because it demands abstract thinking, traditional cognitive behavioral therapy (CBT) has proven difficult for those with ASD; nevertheless, virtual reality (VR) offers a viable substitute. Through controlled exposure to a 360-degree projection of objects they fear, participants in the "Blue Room" gradually gain confidence under the guidance of therapists utilizing cognitive behavioral therapy (CBT) approaches. The majority of participants in two studies—one on adults and the other on children—showed improved functionality in daily life and considerable gains in controlling real-world phobias. The study highlights VR's potential as a formidable tool in ASD therapy, offering a methodical but adaptable way to get over anxieties [34]. Figure 5. Shows the papers distribution considering various categories. This study demonstrates promising advances in autism education through immersive virtual reality (VR). By co-designing a VR version of the Individual Work System (IWS) from the TEACCH® approach with autistic pupils and teachers across the UK, Spain, and Turkey, researchers created a safe and engaging virtual environment tailored to autistic learners, including those with intellectual disabilities. The multi-site trial shows high usability (average SUS score of 85.36) and no significant safety issues, with all 21 participants completing tasks successfully. However, autistic students with intellectual disabilities found the system less feasible, highlighting the need for further adaptations. Overall, the research provides valuable evidence for

VR's potential to support individualized, engaging, and scalable educational interventions for autistic learners [52]. The figure 6(a) and (b) shows the participant and interface of VR.

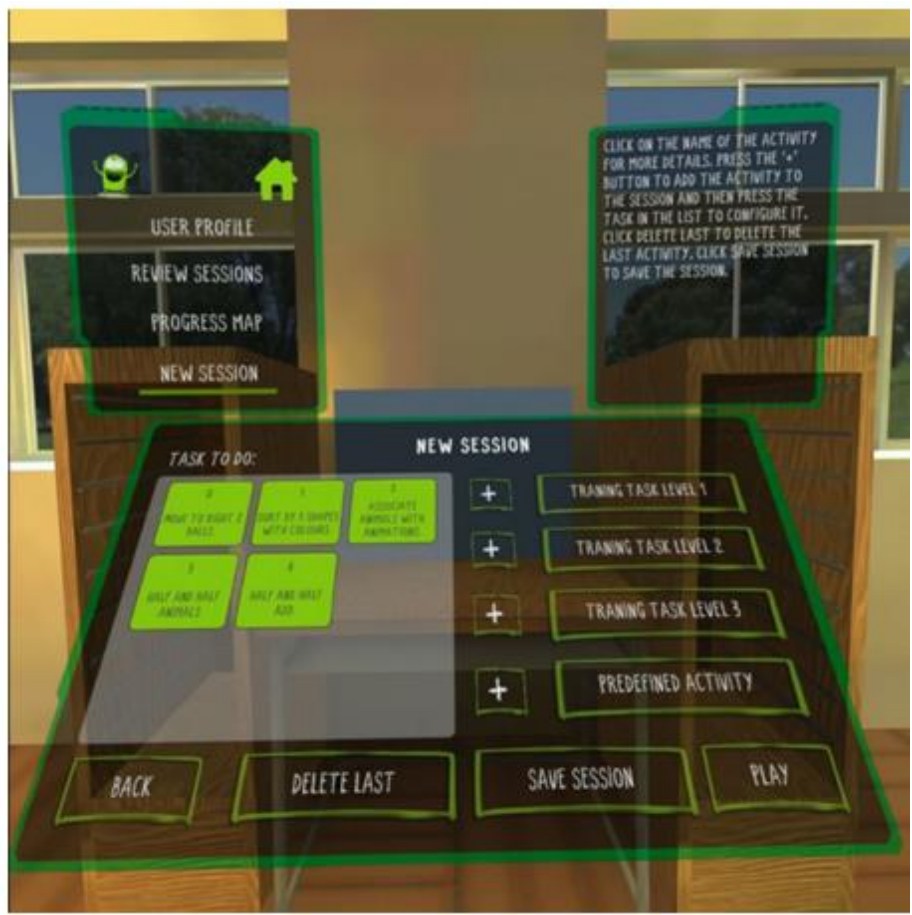

Fig 6(a): VR IWS Interface in study [52]

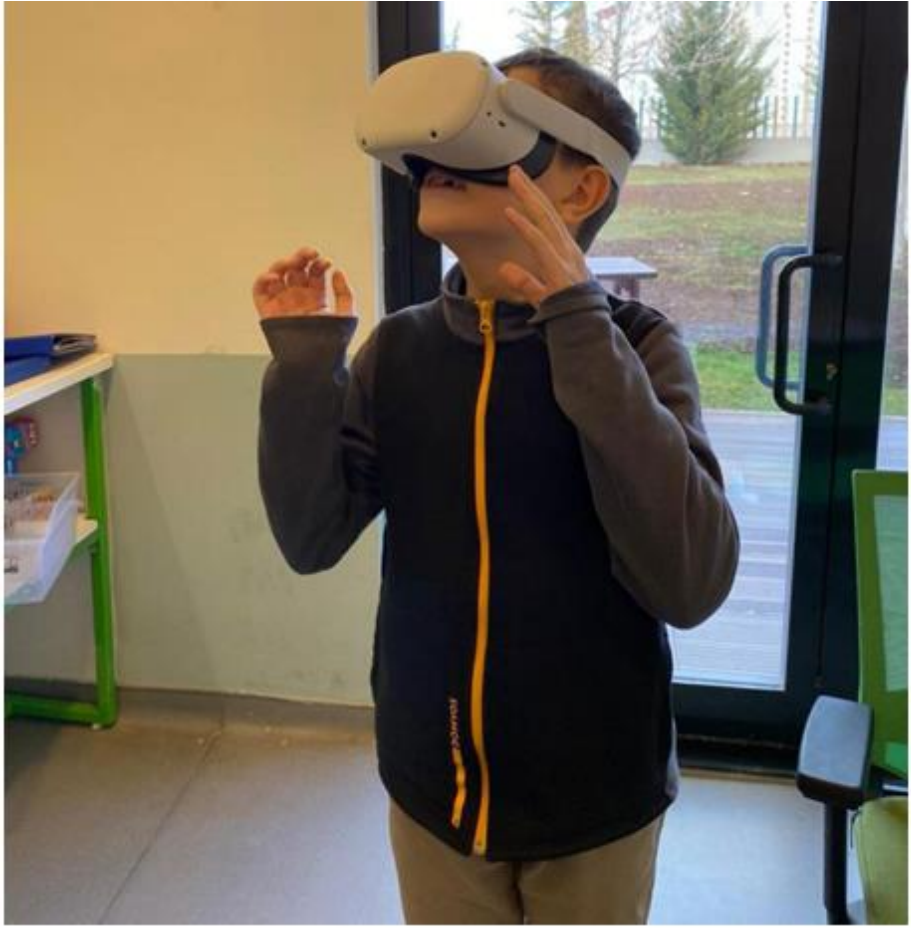

Fig 6 (b): A participant using VR [52]

The study [35] offers a creative and methodical strategy for teaching kids with autism spectrum disorder (ASD) through virtual reality (VR). Compared to traditional approaches, the design saves time and money by using Unity3D to build a virtual environment that mimics real-world circumstances and facilitates immersive learning. The system's usefulness is improved by the inclusion of a website for managing child information and updates, enabling smooth communication between parents, specialists, and the VR environment. In a study Silva et al. (2024) addressed a crucial gap in inclusive education by exploring how Virtual Reality (VR) can support STEM learning for students with Autism Spectrum Disorders (ASD). Using Design Science Research and the Delphi method, the researchers developed and validated a VR artefact tailored to the unique needs and learning styles of these students. Experts strongly endorsed the approach, recognizing its potential to enhance engagement and inclusion, though further refinements are needed for broader effectiveness and the work marked as a significant step toward innovative, accessible STEM education for learners with ASD [25].

The study offers a creative and methodical strategy for teaching kids with autism spectrum disorder (ASD) through virtual reality (VR). Compared to traditional approaches, the design saves time and money by using Unity3D to build a virtual environment that mimics real-world circumstances and facilitates immersive learning. The system's usefulness is improved by the inclusion of a website for managing child information and updates, enabling smooth communication between parents, specialists, and the VR environment. The usability testing result can be seen in figure 6.

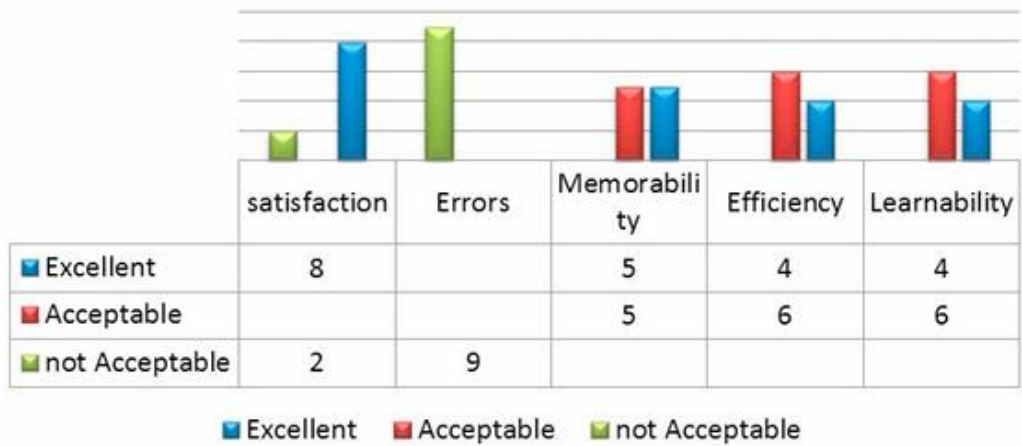

Fig 7. Austistic children usability testing results [35]

Children who participated in testing showed better engagement and adaption to the VR setting, which is encouraging. Test-related difficulties, such getting users acclimated to virtual reality technology, are resolved by educational movies. Upcoming improvements will encompass adding sentence-level interactions, enhancing the realism of the locations, and broadening the VR scenarios to encompass a variety of places such as shopping centers and hospitals. All things considered, the AutiVE system shown in figure 7. has considerable promise as a teaching aid, fostering the growth of kids with ASD and representing a major advancement in the use of virtual reality into teaching methods. A useful tool for behavioral research, virtual reality (VR) systems allow for the construction of immersive 3D settings and precise monitoring of reactions to visual stimuli. Using this technology, researchers can gather data from sizable participant groups outside of conventional labs and investigate intricate theories like interceptive timing behavior with objects traveling in non-standard trajectories. Nevertheless, game developers—not behavioral scientists—are the target audience for current 3D graphics engines like Unity. The Unity Experiment Framework (UXF) was created to make the process of developing VR experiments simpler in order to address this. Using Unity's full potential for stimulus display, UXF offers a programming interface for encoding typical experimental features and changing independent variables [36].

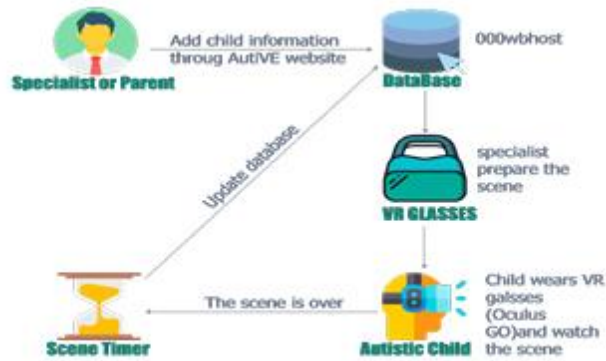

Fig 8. AutiVE system employed in the work [35]

Primarily aimed at helping children with modest learning difficulties develop their socio-emotional skills, augmented reality (AR) has demonstrated potential in improving autism interventions. Approaches that cater to the needs of autistic children with complicated demands and significant learning difficulties are nevertheless required (SLN). In response, "Magic Bubbles"—a multimodal augmented reality environment—was built with input from stakeholders, modified for usage in a day hospital setting, and tested for acceptance and usefulness with children who have SLN and autism. Three main questions were examined in this study. Over the course of three months, seven children with SLN attended at least six weekly sessions. The study used grounded theory to evaluate the children's experiences, combining qualitative and quantitative methodologies. While emphasizing opportunities for future growth, the results also demonstrate that Magic Bubbles is helpful in aiding children with SLN, enhancing their interactions with practitioners, and offering a positive overall experience [37].

With its tremendous impact on recreational and educational activities, virtual reality (VR) technology holds great potential for children with specific educational needs (SEN), especially those with autism spectrum disorders (ASD). For students with ASD, virtual reality (VR) provides a range of tools, including interactive 3D simulations, that can improve social, behavioral, emotional, and cognitive skills as well as everyday chores. The advantages of virtual reality (VR) in education are emphasized in this research review, which also mentions a variety of tools and apps utilized with ASD children. Although there is hope for virtual reality, further study is required to completely comprehend how beneficial it is in learning environments. The paper offers a thorough analysis of recent VR applications along with case studies that show how effective the technology is in helping kids with ASD. It highlights how important it is to have a strong theoretical foundation in order to optimize virtual reality's educational advantages and raise these pupils' standard of living. The review also predicts that in the future, educators will be able to provide more customized virtual reality experiences for students with ASD [38].

This study delves into the application of virtual reality (VR) technologies in the education of children with autism, an area of study that has been ongoing for more than 20 years. In safe and regulated virtual environments, virtual reality (VR) provides immersive "real world" situations that can support social and life skills training. There is now more interest in the possible uses of VR head-mounted displays (HMDs) for autism education due to the recent release of reasonably priced models like Google Cardboard and Oculus Rift. After a thorough analysis of empirical research on VR-HMDs for people with autism spectrum disorders, few studies with different participant characteristics, applications, and technological backgrounds were found. Though the report emphasizes the need for more research to generate strong recommendations for establishing and maintaining VR-based educational techniques, there is some hope over VR's potential benefits. Being the first to assess the data supporting VR-HMD technology in helping people with autism, this literature study is noteworthy [39]. In order to evaluate the impact of immersive Virtual Reality (VR) approaches on cognitive, social, and emotional skills in children and adolescents with Autism Spectrum Disorders (ASD), this research systematically evaluates and meta-analyzes Randomized Controlled Trials (RCTs). The study indicates that, in comparison to control groups, VR therapies significantly improve social skills, emotional skills, and cognitive abilities by examining data from six RCTs. The assessment emphasizes VR's promise as a useful tool for honing these abilities in a flexible and regulated setting, but it also points out issues with accessibility, cost, and customisation. In order to better serve people with ASD, the study recommends conducting additional research to address these issues and improve VR-based therapies [41].

A meta-analysis carefully reviewed and synthesized data from 33 research in order to assess the efficacy of virtual reality (VR) on training and rehabilitation for people with autism spectrum disorder (ASD). Using a wide range of VR and ASD-related search phrases, a thorough search was carried out across several clinical and technical databases, including PubMed, ERIC, Web of Science, PsycINFO, and IEEE. Research on randomized controlled trials and other strong designs that evaluated VR therapies for ASD were the main focus of the rigorous inclusion criteria that were used to choose the studies. Effect sizes were used to assess the data, and subgroups were looked at depending on a variety of factors, including age, the kind of VR technology being used, and the existence of comorbidities. Subgroup meta-regressions were used to examine the influence of continuous moderator variables, and the meta-analysis took into account both formal and non-formal measures of efficacy [42].

In a research study the authors examined how various immersive technologies affected the ability of four to six-year-olds to share physical stickers, walk upon request, and exercise social conformity. The study focused on the children's media character Groover from Sesame Street. Youngsters (N = 52) used Grover on TV or in virtual reality to perform the Simon Says inhibitory control test. When playing Simon Says, children who used VR were less likely than those who used TV to repress a dominant motoric response, which is when they didn't mimic Grover's motions at the right moment. In comparison to the TV condition, more kids in the VR condition approached Grover and shared more stickers with him (among those who shared). Conditions did not differ in terms of children's pleasure of the experience or their level of emotional or physical suffering [44].

The meta-analysis showed that VR training has a large overall effect size (Hedges' g = 0.74), significantly improving a variety of skills in people with ASD. With an effect size of 1.15, the analysis showed that improving daily living skills was the area in which VR interventions were most successful. For social and communication abilities (g = 0.69), emotion control and recognition (g = 0.46), and cognitive skills (g = 0.45), moderate effects were seen. Furthermore, augmented reality-based therapies displayed a positive effect size of 0.92. These results highlight VR's potential as a helpful tool for ASD rehabilitation, but they also highlight the need for additional study to solve standardization and personalization concerns [42].

The ARISP module for teaching and learning activities is observed being implemented, and semi-structured interviews with preschool instructors are part of the research technique. The study looks at how kids engage with augmented reality (AR) technology, how they react to an immersive learning environment, and how they acquire the abilities necessary for digital literacy. The ARISP module considerably raises children's motivation and participation in learning activities, according to the findings. AR technology's interactive features improve problem-solving abilities and encourage a deeper comprehension of STEM subjects. The ARISP module has great potential for improving early childhood education through the combination of innovative technology and conventional teaching methods [46].

A thorough meta-analysis of 134 (quasi-)experimental research on the use of augmented reality (AR) in education from 2012 to 2021 is presented in this article. It classifies how AR affects learning outcomes at three different levels: performance, knowledge and skill, and response. According to the analysis, augmented reality (AR) improves all three outcomes, with performance being most affected. The meta-regression's main finding is that treatment duration has a big impact on how much AR happens. Furthermore, compared to science courses, AR tends to encourage greater learner responses in language and social studies. The study also emphasizes how crucial it is to carefully plan and assess 3D visualization in augmented reality. Ultimately, the article explores the consequences for utilizing AR in the classroom and offers potential avenues for further investigation [47]. The figure 8. Shows the studies comparison in VR and AR. A study carried out on [22] provides valuable insights into how children with autism spectrum disorder (ASD) interact with mobile augmented reality (AR) application interfaces, particularly in coloring apps. Using eye-tracking, the research highlights that ASD children are more drawn to icons and images, which aid task understanding, but can also be distracted by richer, cluttered interfaces. It shows that icon size and placement impact processing speed. The findings emphasize designing AR UIs tailored to ASD characteristics—favoring meaningful visuals, simplicity, minimal distractions, and concise text. This work importantly guides future development of more accessible and engaging AR learning tools for children with autism. The figure 9 shows comparison of VR vs AR interventions and figure 10 shows the distribution of focus categories in pie chart form.

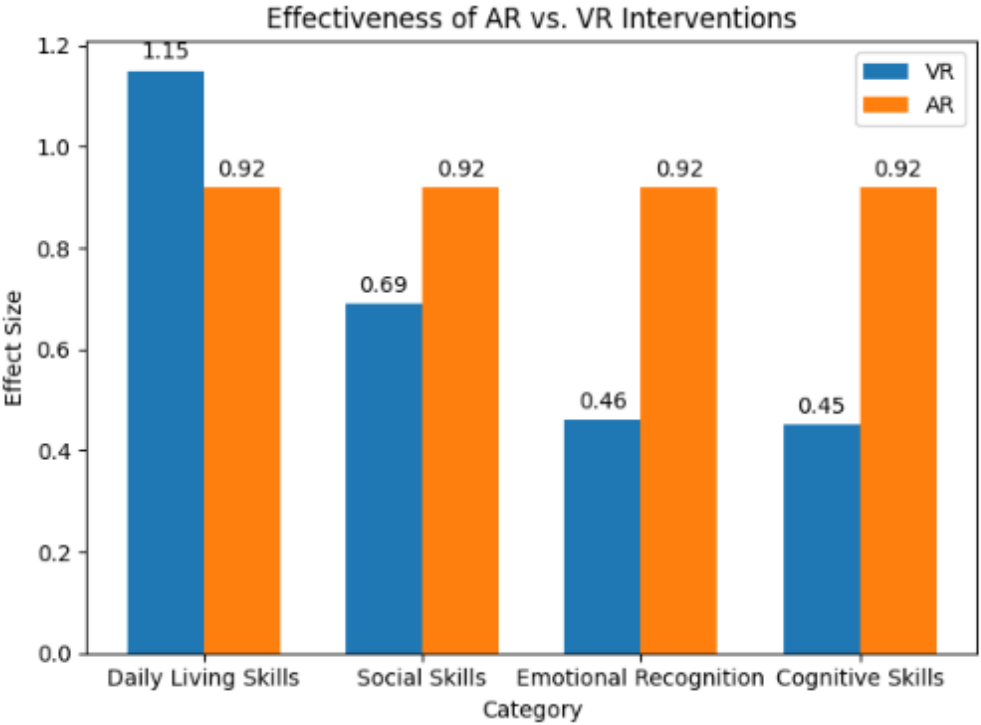

Fig 9. Results from the studies made in AR and VR usage

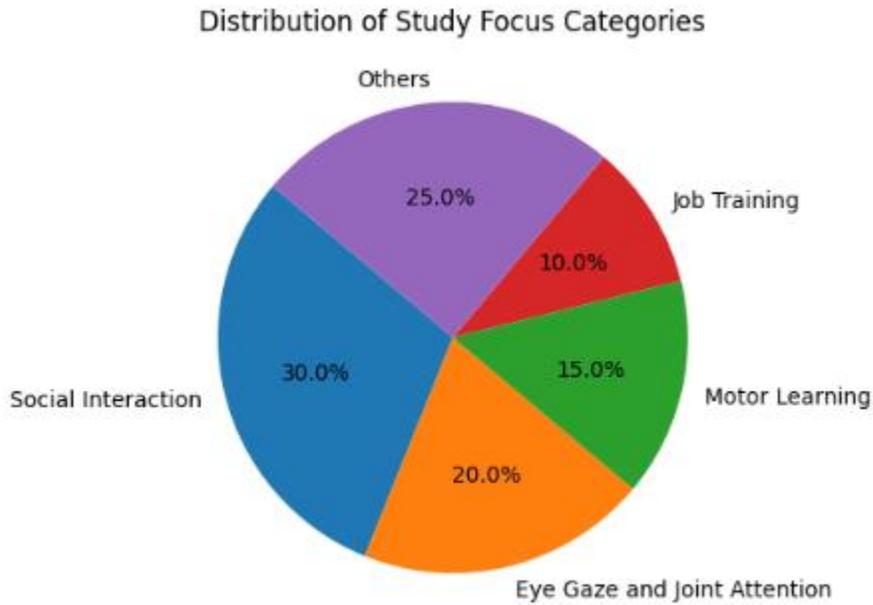

Fig 10. Distributions of study focus categories.

The distribution of research as seen in chart in figure 9. Shows target categories for virtual reality (VR) therapies for children with autism spectrum disorder (ASD) is depicted in the pie chart. It displays the distribution of research across a number of topic areas, such as joint attention and eye gazing, motor learning, social interaction, job training, and others. The graphic shows that social contact is prominent in VR treatments, with a large amount of study devoted to this topic. While job training and other categories make up smaller chunks, eye gazing and joint attention, as well as motor learning, also receive considerable focus. With a distinct focus on improving social skills and engagement through VR technology, this distribution demonstrates the wide range of applications and interests within the sector. According to the early results [44], VR may cause distinct cognitive and social reactions in children than less immersive technologies.

3.2 AI, social robots and Large Language Models (LLMs) in ASD

The advent of artificial intelligence and LLMs, such as GPT and BERT, has opened new avenues for supporting communication in children with ASD. These models can generate natural language responses, assist in language learning, and provide real-time feedback during interactions. Research by Mishra et al. (2024) [13]. In a most recent study [29] Liu Lan et al. (2025) presents the Public Health-Driven Transformer (PHDT), a new AI model that combines cutting-edge machine learning and public health concepts to improve social skills in kids with ASD. For more captivating social learning, PHDT provides real-time, adaptive feedback based on multi-modal data, such as text, audio, and face cues. According to experimental findings, PHDT considerably enhances social skill acquisition, engagement, and retention when compared to conventional approaches. This study demonstrates how AI-powered, public health-focused solutions have the potential to revolutionize intervention strategies and results by offering children with ASD individualized, easily accessible support. A novel and insightful approach to autism diagnosis by Stanley et al. (2025) using large language models (LLMs) to decode clinicians' diagnostic reasoning from real-world health records. Rather than relying on genome-wide data or imaging—which have yielded limited gains—it taps into the clinical intuition documented in free-text reports. Impressively, the model not only distinguishes confirmed from suspected autism cases but also explains its decisions by highlighting the most influential sentences. Notably, the findings suggest that clinicians' judgments hinge more on behaviors like stereotyped movements and special interests than on social deficits traditionally emphasized in DSM-5, pointing to potential revisions in diagnostic frameworks. This work shows the AI emerging impact on redefining autism diagnosis by translating clinicians' nuanced observations into measurable patterns, offering both explainability and fresh perspectives that could reshape diagnostic criteria and improve patient care.

LLMs can also be integrated into AR environments to enhance the interactivity and relevance of the content presented to children with ASD. For instance, LLMs can provide explanations or prompts during AR-based learning sessions, helping children better understand the tasks they are engaged in. The potential for LLMs to be used as assistive technologies for communication is also explored by Woolsey et al. (2024) [15], who highlight the importance of designing these models with accessibility in mind to ensure they are effective for children with varying levels of ASD. The applications of large language models (LLMs) and other therapies in the emotional and developmental contexts of children and individuals with autism are summarized in this review of recent research. A notable study presented ChaCha, a chatbot as shown in figure 11designed to help kids express their emotions. Twenty participants, ages eight to twelve, were included in the exploratory study. It was shown that ChaCha successfully encouraged kids to communicate intimate details and feelings, which made the chatbot seem like a close friend. The study demonstrated the potential of this LLM-powered chatbot to support emotional conversations with children. Notably, 19 out of 20 participants indicated they would be open to engaging with ChaCha for longer periods in the future, although their preferred frequency of interactions varied and sustained use of ChaCha could enhance the advantages of employing LLMs, such as fostering rapport with children [26].

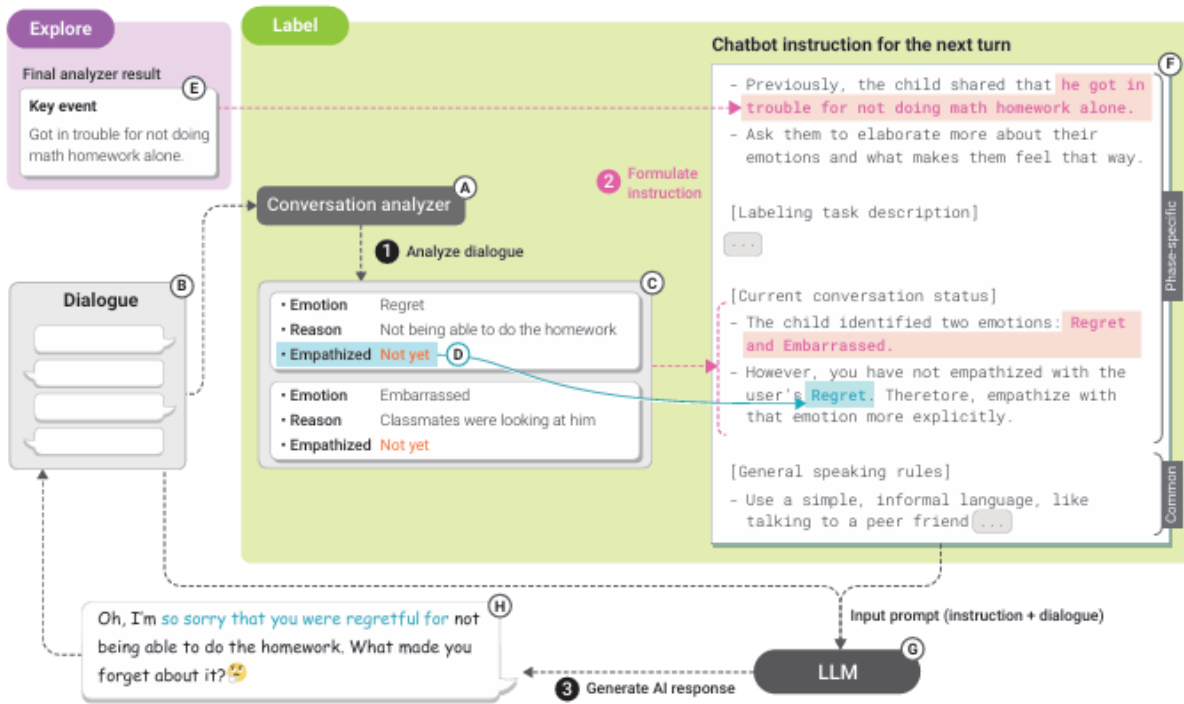

Fig 11: ChaCha chatbot functionalities [26]

This suggests that LLMs have the ability to foster nurturing environments for kids, assisting them in navigating the often difficult process of emotional communication because of their immature language skills. The advent of generative AI tools in education, especially LLMs like ChatGPT, has forced schools to reevaluate their writing curricula [27]. Although there is a lot of interest in using LLMs to improve kids' writing skills, there isn't much research available at the moment. Since LLMs were originally intended for adult users, there are still questions over whether or not they are suitable for younger users, despite the fact that their early use suggests they can offer helpful writing support.

The contribution of LLMs to raising children's health literacy is a crucial topic of research. A study evaluated various LLMs' responses to questions regarding pediatric medical issues, with an emphasis on the outputs' readability [28]. The outcomes showed that LLMs could provide responses that were appropriate for different reading levels, indicating their potential as instruments to support kids in understanding complicated health information. The study did note, however, that creating content that is suitable for younger audiences presents certain difficulties, indicating the need for additional LLM capability improvement.

Adults with autism who have social communication issues can also benefit from LLMs. A study looked at how these people used an LLM chatbot to get help with social interactions at work [29]. The fact that participants clearly preferred the chatbot to a human confederate shows how valuable LLMs are at offering a platform where people can ask for assistance without feeling judged. There is a conflict between the requirement for normative guidance from subject-matter experts and user happiness, as seen by concerns expressed on the caliber of counsel provided by the LLM. This work [16] highlights the promise of Large Language Models (LLMs) as personalized educational tools for children with Autism Spectrum Disorder (ASD), addressing their unique social and sensory challenges. The educational method powered by AI creates a relaxed learning atmosphere that reduces obstacles for children with ASD, while giving parents and teachers immediate information to customize education. It fosters the development of social skills via interactive simulations and encourages lasting societal advantages by enabling autistic individuals to make meaningful contributions in various areas.

In a research by Bernabei et al. (2023) three main studies are done: the caliber of essays written by students who use LLMs, the efficacy of LLM detection systems, and the views of students regarding the use of LLMs in the classroom. The study indicated places for improvement for both students and educators, but it also demonstrated that essays written with LLM support were of good quality. Thirteen LLM detectors were tested, and none of them worked well enough to support the conclusion against outlawing LLMs. Furthermore, according to recognized technological acceptance models, a questionnaire revealing student evaluations of the usefulness and acceptance of LLMs was good. The study emphasizes the necessity of more investigation and modifications to teaching methods, particularly with regard to the incorporation of LLMs into classroom settings [50]. Many studies are there on therapies for kids with autism spectrum disorder (ASD) that have demonstrated how successful evidence-based methods like Applied Behavior Analysis (ABA) are at lowering stereotyped behaviors [30]. Significant improvements were observed in the study among the patients who received interventions, highlighting the significance of early therapy and family participation. Future research should examine broader factors influencing treatment outcomes, as the data also indicated gender variations in response to interventions.

These studies show the ways that LLMs and organized interventions can promote emotional communication, improve educational practices, raise health literacy, and assist people with autism, but they also point up areas that need more investigation and improvement. The figure 10 highlights the research areas of LLMs in enhancing the communication for ASD.

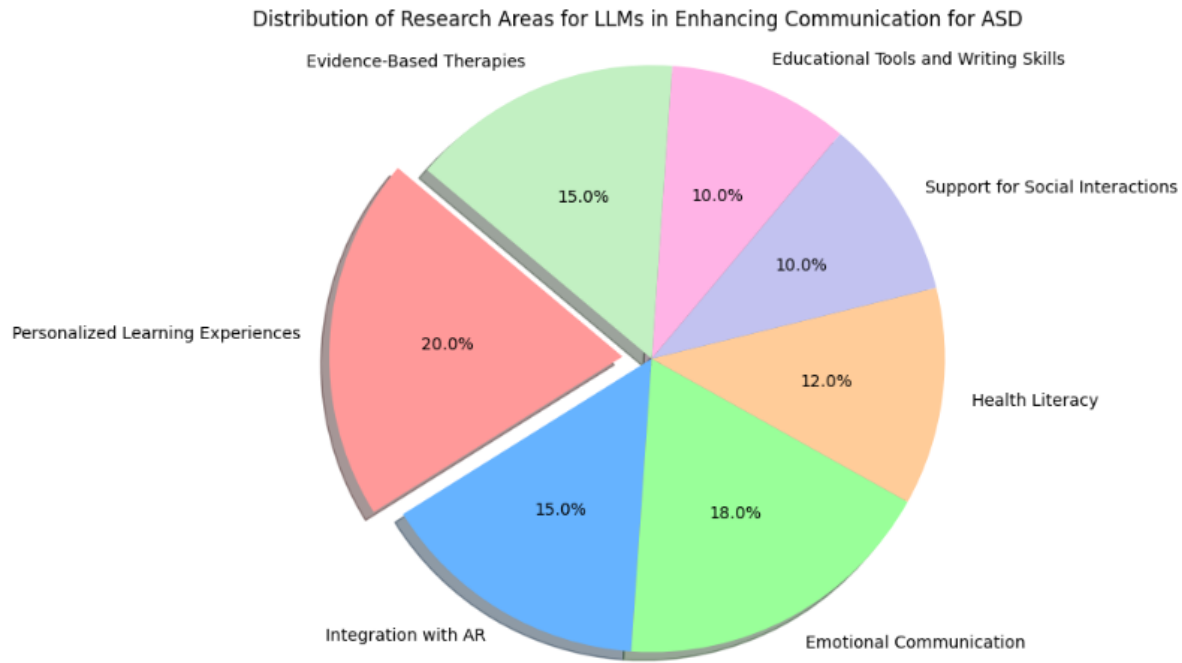

Fig 12. Distribution of research areas of LLM's for ASD

The distribution of study as shown in figure 12 emphasis areas for Large Language Models (LLMs) in improving communication for kids with Autism Spectrum Disorder (ASD) is depicted in the pie chart. The largest part, which stands for Personalized Learning Experiences, emphasizes how much research is focused on customizing LLMs to modify content according to each student's level of language competency and communication requirements. Following integration with AR, there is a clear focus on enhancing interactive learning sessions by integrating LLMs with AR. '

A lot of focus is also paid to emotional communication, which highlights initiatives to employ LLMs to support kids in expressing and comprehending their feelings. Another important area of research is health literacy, where LLMs are assessed for their capacity to make medical information understandable to younger audiences. According to Support for Social Interactions, LLMs are being investigated as a potential tool to support social skills in adults and children with ASD. The chart also shows interest in using LLMs to enhance writing skills in school settings by include Writing Skills and school Tools. Finally, research on combining LLMs with well-established therapy approaches is highlighted in Evidence-Based Therapies. This distribution emphasizes a thorough strategy for applying LLMs to different facets of therapy, education, and communication for people with ASD. Even while large language models (LLMs) acquire four or five orders of magnitude more language data than human children, they nonetheless exhibit fascinating emergent patterns [48]. This research examines the application of social robots, particularly the NAO robot, to enhance autism diagnosis by concentrating on turn-taking interactions in preschoolers. Through the comparison of interactions between autistic children and typically developing peers during two robot-led games, the study uncovered notable deficits in turn-taking for children with autism. Through an observational rating scale and statistical analysis, the research illustrates the capability of interactive robot games as effective instruments for identifying social communication difficulties. The results back the creation of robotic technologies to assist in prompt and precise autism diagnosis [53]. The work by Kumazaki et al., (2020) highlights the careful consideration required in choosing robot types—from simple or animal-like robots that facilitate easy interaction, to android robots that may better support generalization into real-life contexts. The review underscores how robot features like appearance, movement, and demeanor, along with user characteristics such as age, sex, and IQ, influence engagement and effectiveness. While the potential for robotics in ASD interventions is promising, the paper stresses that few empirical studies exist so far, calling for further research to optimize robotic designs and applications in this field [59].

Mishra et al., (2024) study advanced robotic interventions for children with Autism Spectrum Disorder (ASD) by moving beyond rigid, pre-scripted interactions to a more autonomous, perspective-taking approach. Leveraging large language models (GPT-2 and BART), the robot dynamically generates social scenarios, prompts, and positive reinforcements, embodying multiple teaching roles. The pipeline was validated through expert simulations and evaluated using established measures (NASA TLX, GodSpeed, BERTScore), showing improved interaction quality without increasing user cognitive or physical load. Experts also perceived the robot as safe, likable, and reliable. Overall, this work demonstrates a promising step toward more flexible, engaging robotic therapies for ASD [56]. The NAO robot for teaching of children is depicted through figure 13.

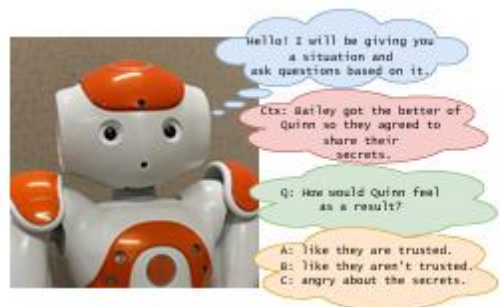

Fig 13. NAO robot for teaching children with ASD [56]

### 3.3 UI/UX Design in ASD-Focused Applications

Effective UI/UX design is critical in ensuring that AR and LLM-based interventions are accessible, engaging, and beneficial for children with ASD. The user interface must be intuitive, with clear instructions and minimal distractions, to accommodate the cognitive and sensory sensitivities common in children with ASD. Studies by Lian et al. (2023) and Putri et al. (2025) emphasize the need for UI/UX designs that are tailored to the unique needs of children with ASD, including the use of calming colors, simple layouts, and easy-to-navigate menus.

Moreover, eye-tracking studies, such as those conducted by Lian et al. (2023) [22], have shown that children with ASD can benefit from interfaces that adapt to their gaze patterns, reducing the cognitive load and enhancing their engagement with the content. UI/UX designs that incorporate feedback mechanisms, such as rewards or progress indicators, can also motivate children to engage more consistently with the interventions.

A study by Putri et al. (2025) presents SemaiSelaras, an innovative adaptive learning app designed for children with Autism Spectrum Disorder (ASD). Uniquely combining the Applied Behavior Analysis (ABA) approach with Design Thinking (DT) methodology, the work effectively addresses the specific learning needs of children with ASD, focusing on teaching essential adaptive living skills. The integration of technologies like OCR, digital storyboards, audio discrimination, and video-based learning strengthens the app's versatility [18]. The reported System Usability Scale (SUS) score of 86.5 indicates excellent usability and user acceptance. Overall, the study makes a valuable and significant enhancements to the field of inclusive educational technology, demonstrating how ICT can enhance accessibility and skill development for children with ASD. Pavlov et al. (2014) explores the design of accessible user interfaces for individuals with Autism Spectrum Disorders (ASD), focusing on the Open Book reading assistive tool. It combines insights from prior research on reading comprehension in ASD with feedback from users and professionals to define UI requirements [9]. Ismali et al. (2021) reviewed the role of smartphone-based emotion-learning applications in supporting individuals with Autism Spectrum Disorder (ASD) and their caregivers. It emphasizes the importance of UI/UX design in shaping learning and emotional expression, highlighting how differences in app functionality impact user experience. The paper also explores **the** association between colors and emotions, noting how color can aid in emotional recognition and communication for users with ASD. While largely a literature review, the work provides valuable insights into the intersection of design, technology, and emotional learning in ASD-focused applications, underscoring the importance of personalized and intuitive interfaces. Valencia et al. (2025) addressed the critical gap in UX evaluation methods tailored for adults with Autism Spectrum Disorder (ASD)**.** It introduces and validates a three-stage UX/ASD methodology designed specifically to assess user experience beyond generic usability tools. Using the PlanTEA application—an app helping ASD users plan medical visits—as a case study, the authors demonstrate how their approach yields valuable insights to enhance satisfaction and usability. The work effectively integrates perspectives from UX experts, ASD professionals, and potential users, highlighting the need for specialized evaluation techniques. Overall, it makes a significant contribution to **inclusive UX research**, offering a practical framework for assessing and improving digital tools for adults with ASD [10].

In a study by Kamaruzaman et al. 2016 [51] explores the growing role of touchscreen-assistive technology in supporting children with autism, especially in learning basic numeracy. Through a participatory design approach involving autistic children, the researchers developed the Touchscreen-Assisted Learning Numeracy App (TaLNA) as shown in figure 5. TaLNA combines interactive animations along with a user-friendly interface to enhance number recognition and calculation skills. It complements traditional tools like cue cards and offers a practical resource for parents and educators to enhance the engagement and learning outcomes in a more accessible and stimulating way. The gaming or similar fun approach can help learning as well as Apps like TaLNA as shown in figure 14. benefit children with autism by using colorful visuals and interactive features to make learning more engaging and easier to understand. They help to build and imporve focus, support step-by-step learning of skills like counting and number recognition, and offer a fun alternative to traditional tools like cue cards. Such apps also empower parents and teachers to teach more effectively, helping children build confidence and independence in learning.

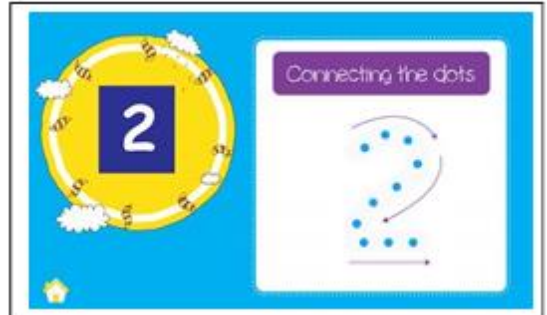

Fig 14: Connecting the dots in TaLNA [51]

This study examines the complex interrelationships between User Interface (UI) and User Experience (UX) design, emphasizing the differences and similarities between the two fields. It highlights how crucial user-centered design is becoming in the rapidly changing digital environment. The study explores how new developments, mainly in artificial intelligence (AI), are influencing customer satisfaction overall by reshaping human-computer interaction (HCI) and user experience (UX). Through an examination of these developments, the study provides insightful information on the direction digital product design will take and how the customer experience will change in the future [32].

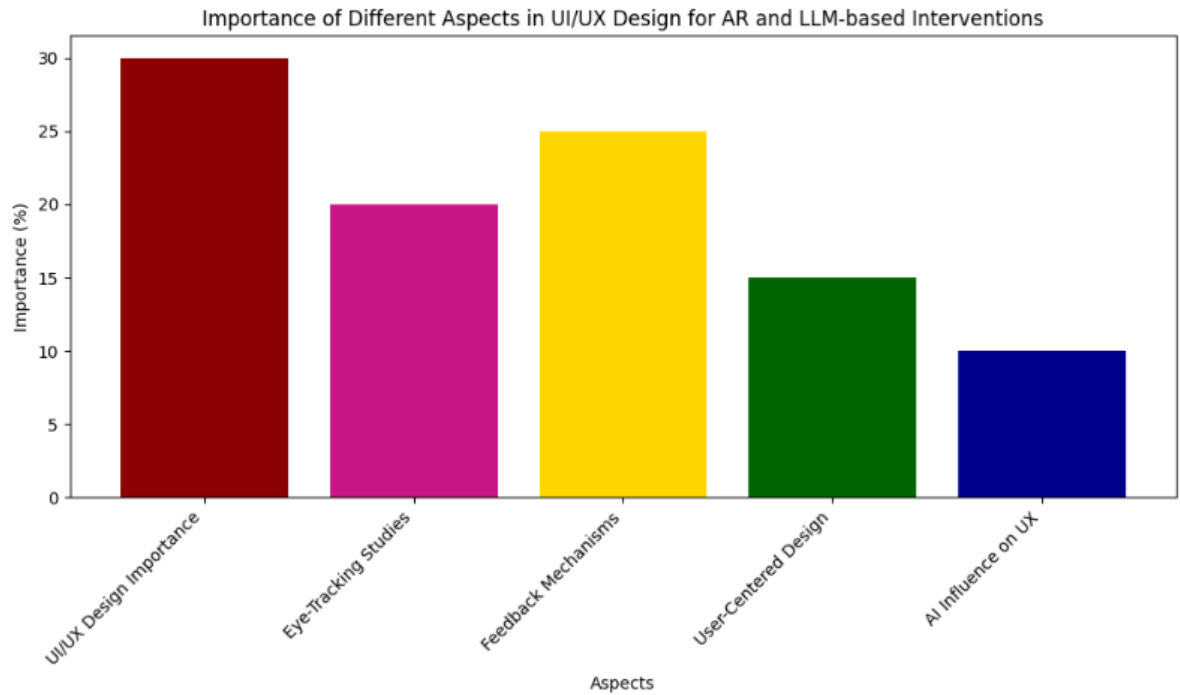

Fig 15. Importance of different aspects In UI/UX design for AR and LLM-based interventions

The bar graph in figure 15.shows how important different UI/UX design elements are for augmented reality (AR) and language learning (LLM) interventions that are suited for kids with autism spectrum disorder (ASD). To accentuate their distinct functions, dark hues are used to highlight each bar, which stands for a distinct design aspect. The most important feature, "UI/UX Design Importance," is highlighted in dark red and indicates that developing an interface that is both user-friendly and adaptive is essential to effectively engaging children with ASD. The value of creating interfaces that adjust to gaze patterns is highlighted by the "Eye-Tracking Studies," which are displayed in dark pink. This reduces cognitive burden and improves engagement.

"Feedback Mechanisms," which are indicated in dark yellow, stress how crucial it is to include elements like progress markers and prizes in order to increase motivation. Dark green is used to symbolize "User-Centered Design," which emphasizes the importance of creating user interfaces that are oriented around their requirements and preferences. Lastly, "AI Influence on UX," highlighted in dark blue, highlights how artificial intelligence is increasingly influencing how users interact and experience the world. Overall, the graphic highlights the necessity of well considered UI/UX design that is adapted to the unique requirements of kids with ASD in order to increase efficacy and engagement. The improved learning scheme is shown in figure 12. A comparison of some of major research and review works is given in table 1. Which further shows the significance of such technologies in the field of immersive education for children. Figure 16 depicts an improved learning scheme for children.

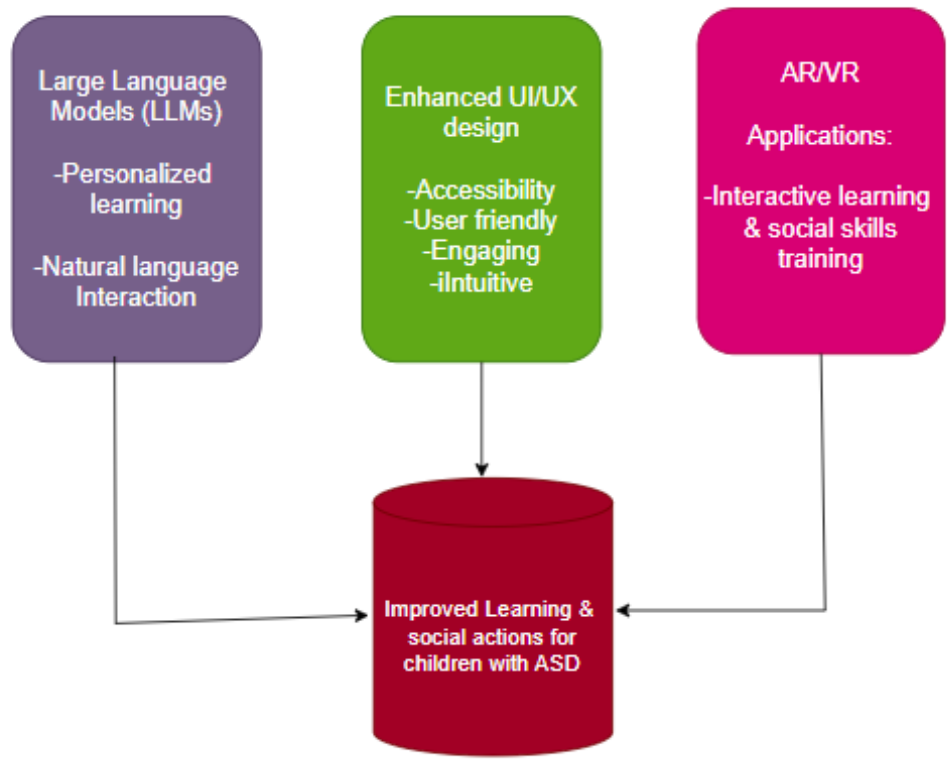

Fig 16. Improved Learning scheme for ASD affected children

Table 1. Previous literature and the findings

| Study | Year | Focus Area | Conclusions |
|---|---|---|---|
| Wiggins et al. | 2023 | AR/VR in ASD | AR/VR enhances social interactions and learning in children with ASD by providing immersive experiences. |
| Maskey et al. | 2019 | Virtual Reality (VR) in Education | VRE combined with CBT is a feasible, acceptable, and partially effective treatment for specific phobias in autistic people. |

| Lian et al. | 2023 | Evaluation of UI of a mobile augmented reality coloring application | Emphasizes on designing AR UIs tailored to ASD characteristics—favoring meaningful visuals, simplicity, minimal distractions, and concise text. |
|---|---|---|---|
| Seo et al. | 2024 | Chatbot ChaCha focused on ASD | LLM chatbot's effectiveness in facilitating emotional communication. |
| Ramos Aguiar et al. | 2023 | Gamification for Autistic People with Tangible Interfaces | Using the developed app enhanced performance |
| Mishra et al. | 2024 | robot autonomy in ASD therapy by using large language models | GPT-2 + BART-based robotic intervention effectively supported perspective-taking teaching |
| Wang et al. | 2025 | Gamification in ASD reviewed | Gamification motivates learning and improves social interaction and communication in ASD children through rewards and challenges. |
| Bernabei et al. | 2023 | Natural Language Processing (NLP) for children education | NLP helps in understanding and generating language that children find easier to engage with. |
| Khowaja et al. | 2020 | Reviewed use of augmented reality (AR) to improve various skills. | It is not possible to say how well AR works for teaching or learning abilities connected to ASD based on what is already known. |
| You et al. | 2023 | Cognitive Behavioral Therapy (CBT) | CBT combined with digital tools offers an effective approach to managing ASD symptoms. |
| Huber et al. | 2024 | Educational game in ASD | Cognitive functions and behavior development for children with game. |
| Bhatt et al. | 2014 | Educational Games for ASD | Educational games can be designed to improve specific cognitive functions in ASD children. |
| Abdallah et al. | 2017 | Brain-Computer Interfaces (BCIs) in ASD | BCIs offer new ways for non-verbal ASD children to |

| | | | communicate through brain signals. |
|---|---|---|---|
| Reed et al. | 2021 | Social Media in ASD Support | Social media platforms can provide support networks for ASD individuals and their families. |
| Arent et al. | 2022 | Social Robots in ASD Therapy | Social robots can effectively model and reinforce appropriate social behaviors in ASD children. |
| Lan et al. | 2025 | Public Health-Driven Transformer (PHDT), an AI model that uses multi-modal data (text, audio, facial cues) to deliver adaptive, real-time social skills training for children with ASD. | PHDT outperformed traditional methods by significantly improving engagement, retention, and social skill acquisition in children with autism. |

## 4. Synthesis of Findings:

Research indicates that both Virtual Reality (VR) and Augmented Reality (AR) greatly enhance social, cognitive, and emotional abilities in children with ASD [11]. However, VR typically shows more pronounced effects in immersive social scenarios (such as joint attention and phobia therapy), while AR tends to excel in fostering participation in educational settings like STEM activities and language exercises. Meta-analyses reveal significant effect sizes for VR in daily living skills and moderate effects for AR in educational outcomes, indicating that VR might be more effective for immersive social training, while AR provides practical advantages in classroom learning and targeted interventions. AI based gaming and education applications can be helpful to enhance cognitive and knowledge for children.

Research on robotics [53, 56] in autism interventions shows promising results, showing that social robots like NAO can effectively identify social communication challenges, such as difficulties with turn-taking, in children with ASD. Robotics interventions have progressed from rigid, pre-scripted interactions to more sophisticated systems powered by large language models (LLMs) [13, 14], allowing robots to dynamically generate social scenarios, deliver adaptive prompts, and provide positive reinforcement. These advancements enable robots to take on multiple teaching roles, enhancing engagement without increasing cognitive or physical strain for children. Robotics systems serve both diagnostic and therapeutic purposes, helping assess social skills while also fostering improvements in social, emotional, and cognitive development. However, further research is needed to evaluate the long-term impacts, scalability, and cost-effectiveness of integrating robotics into widespread autism interventions.

**4.1 Theoretical Implications**:

These findings matches well with Social Cognitive Theory, emphasizing the role of observational learning, modeling, as well as reinforcement in behavior change. Technologies like VR and AR create safe, controlled environments where children with ASD can practice social scenarios, receive immediate feedback, and model appropriate behaviors, enhancing self-efficacy and skill gain for children. Also, the integration of AI and LLMs [13, 14, 15] helps on gaining of personalized interventions, supporting individualized learning paths consistent with constructivist and socio-cognitive learning frameworks.

### 4.2 Practical Recommendations:

4.2.1 Design Guidelines for ASD-Focused Tech:

  i. Use calming colors and simple layouts to minimize sensory overload.
  ii. Incorporate clear instructions, minimal text, and meaningful visuals tailored to ASD cognitive profiles.
  iii. Enable gaze-adaptive interfaces (e.g., eye-tracking) to reduce cognitive load.
  iv. Implement feedback mechanisms like rewards and progress indicators to sustain engagement.
  v. Integrate customizable settings to match individual sensory and learning needs.

4.2.2 Policy Suggestions for Implementation in Schools:

  i. Encourage pilot programs using VR/AR tools for social and emotional learning in special education.
  ii. Provide training for educators on integrating immersive technologies into ASD interventions.
  iii. Develop standards for data privacy and ethical use of AI and VR tools with vulnerable populations.
  iv. Allocate funding for research into cost-effective, scalable solutions for ASD-focused technology.

### 4.3 Limitations:

  i. Many studies involve small sample sizes, limiting statistical power and generalizability.
  ii. Significant heterogeneity exists across studies in terms of participant age, ASD severity, and intervention types.
  iii. Few long-term follow-ups are available, leaving uncertainty about sustained impacts.
  iv. Publication bias may inflate reported effectiveness, as positive results are more likely to be published.
  v. Technological disparities (e.g., differences in hardware or software) complicate cross-study comparisons.
  vi. Clear lack of use of robotics and AI-methods focused on recreational and educational focus for ASD affected.

**4.4 Future Directions**:

i. Longitudinal Studies**:** Needed to evaluate the durability of VR/AR and AI interventions over months or years, especially their impact on real-world functioning. This can be done with survey or case studies on participants and analyzing the data on various kind of ASD affected children.

ii. Hybrid Interventions: Combining modalities such as AR with social robots or LLM-based conversational agents could enhance personalization and engagement. This cross functionality technology development can enhance learning and potential of development in education, adapting the needed cognitive abilities for children.

iii. Cost-Effectiveness Analysis: Such cost-effective analyses are essential for steering educational policy and identifying which interventions yield the most favorable results in relation to the resources invested, particularly for broad implementation in schools or clinics. Affordable alternatives can be explored for children from disadvantaged or privileged groups. Technologies such as VR implementation are expensive, particularly robotic systems that impose a significant financial strain. This must be addressed by discovering affordable approaches centered on creating an educational environment for children with ASD

5. **Conclusion**

The review highlights the noteworthy possibilities of incorporating Large Language Models (LLMs), Augmented Reality (AR), and User Interface/User Experience (UI/UX) design into therapies for children diagnosed with Autism Spectrum Disorder (ASD). By offering immersive and interactive environments that are tailored to the specific needs of kids with ASD, augmented reality (AR) has shown itself to be a useful tool for improving social interaction and learning. Language learning and communication can benefit greatly from the assistance of LLMs, who can design customized and flexible learning programs. It is impossible to exaggerate the significance of good UI/UX design since it guarantees that these interventions are approachable, interesting, and catered to the cognitive and sensory needs of kids with ASD.

Notwithstanding the encouraging results, there are still issues to be resolved, such as the requirement for content personalization, enhancements to accessibility, and improved technology integration. Future studies should focus on resolving these problems and investigating ways to optimize UI/UX design, AR, and LLMs to deliver more tailored and efficient treatments. The quality of life and developmental outcomes for kids with ASD could be greatly enhanced by further developments in these fields.

**Declaration of AI/ LLM tools use:**

The authors declare that they used LLM tools to minimize grammar errors and writing improvements.

**Declaration of competing interests:**

The authors declare that there are no known competing interests

**Ethics Statement**

This research did not involve human participants or animals and therefore did not require ethical approval. The study was conducted in accordance with Elsevier's publishing ethics guidelines. All data analyzed were obtained from publicly available sources, and appropriate attribution has been provided to all referenced works.